\documentclass[prb,superscriptaddress,twocolumn,showpacs,preprintnumbers,amsmath,amssymb,reprint]{revtex4-1}

\usepackage{graphicx}
\usepackage{amsmath}
\usepackage{amssymb}
\usepackage{color}
\usepackage{dcolumn}
\usepackage{epsfig}
\usepackage{bm}
\usepackage{subfigure}
\usepackage[urlcolor=blue]{hyperref}
\hypersetup{backref, colorlinks=true, linkcolor=blue, citecolor=blue}

\begin{document}

\title{Chain length effects of $p$-oligophenyls with comparison of benzene by Raman scattering}

\author{Kai Zhang}
\affiliation{Key Laboratory of Materials Physics, Institute of Solid State Physics, Chinese Academy of Sciences, Hefei 230031, China}
\affiliation{University of Science and Technology of China, Hefei 230026, China}
\affiliation{Center for High Pressure Science and Technology Advanced Research, Shanghai 201203, China}

\author{Xiao-Jia Chen}
\email{xjchen@hpstar.ac.cn}
\affiliation{Center for High Pressure Science and Technology Advanced Research, Shanghai 201203, China}
%\affiliation{Key Laboratory of Materials Physics, Institute of Solid State Physics, Chinese Academy of Sciences, Hefei 230031, China}

\date{\today}

\begin{abstract}
Raman scattering measurements are performed on benzene and a number of $p$-oligophenyls including biphenyl, $p$-terphenyl, $p$-quaterphenyl, $p$-quinquephenyl, and $p$-sexiphenyl at ambient conditions. The vibrational modes of the intra- and intermolecular terms in these materials are analyzed and compared. Chain length effects on the vibrational properties are examined for the C-H in-plane bending mode and the inter-ring C-C stretching mode at around 1200 cm$^{-1}$ and 1280 cm$^{-1}$, respectively, and the C-C stretching modes at around 1600 cm$^{-1}$. The complex and fluctuating properties of these modes result in an imprecise estimation of the chain length of these molecules. Meanwhile, the obtained ratio of the intensities of the 1200 cm$^{-1}$ mode and 1280 cm$^{-1}$ mode is sensitive to the applied lasers. A librational motion mode with the lowest energy is found to have a monotonous change with the increase in the chain length. This mode is simply relevant to the $c$ axis of the unit cell. Such an obvious trend makes it a better indicator for determining the chain length effects on the physical and chemical properties in these molecules.
\end{abstract}


\maketitle

\section{Introduction}

Conjugated polymers have attracted more interests as the novel properties resulted from the delocalized $\pi$ electrons. During the last few decades, they have been applied in a big field of scientific research.\cite{Green,Tang} Among the polymers, poly($para$-phenylene) (PPP) attracts attention for their versatile properties. A blue light-emitting device was produced with PPP thin layers.\cite{Grem,Grem-1} The oligomeric lower members of PPP formed between the rings in the polymerization of benzene (B) were referred to by Kern and Wirth\cite{kern} as $p$-oligophenylenes. These compounds are now named $p$-oligophenyls (POPs) in which the prefix for the number of linked rings (ter-, quater-, quinque-, sexi-, septi-, octi-, novi-, deci-, ere) is placed in front of the radical name phenyl. Among them, $p$-terphenyl (P3P) in solution is made for a highly efficient ultraviolet laser dye.\cite{Cecco} $p$-Quaterphenyl (P4P) and $p$-sexiphenyl (P6P) are used for light emitting diodes.\cite{Hwang,Koch} Meanwhile, organic $\pi$-conjugated POPs are also candidates for superconductors or good thermoelectric materials. It has been reported\cite{Havin,Ivory,Shack} that the conductivity of POPs increases by several orders of magnitude by increasing the chain length of molecules after strong electron doping. Recently, superconductivity was also found in potassium-doped P3P,\cite{Ren-1,Ren-2,Ren-3} P4P,\cite{Yan} P5P,\cite{Huang} and 2,2$'$-bipyridine,\cite{Zhang} an isologue of P2P. These findings are extremely attractive for exploring higher temperature superconductors in organic materials. The significant increase of conductivity also makes these materials potentials for thermoelectric applications.\cite{Dubey,Xuan,Bubno} To improve the performance of these materials, it is important to study their structural characteristics and related properties.

The structure of crystalline POPs was investigated at ambient conditions by X-ray and neutron diffraction experiments.\cite{Charb,Baudo,Baudo-1,Delug} The results show that, unlike the orthorhombic structure of crystalline benzene at low temperature, all the POPs at ambient conditions exhibit a monoclinic structure with the $P$2$_1$/$a$ space group.\cite{Charb,Baudo,Baudo-1,Delug} They almost have the same length along the $a$ and $b$ directions, whereas the c axis increases with increasing the chain length because of the growing length of the molecules that are approximately parallel to the $c$ axis. X-ray diffraction (XRD) data indicated that the phenyl rings in one molecule all have strong thermal librations around the long molecular axes. However, the conformation of the individual molecules is planar on average. There are two different types of forces to make the torsional angle. The $\pi$-conjugation between phenyl rings and between neighboring molecules tends to stabilize the planar conformer. The other force is the steric repulsion between neighboring $ortho$-hydrogen atoms which favours twisting the rings.\cite{Carre,Soto} Raman spectroscopy is a powerful tool to detect the dynamical processes. The Raman spectra of benzene were measured as early as the 1930s.\cite{Wilso,Angus,Angus-1} Later, the phase diagram of solid benzene at high pressures was completed.\cite{Ciabi,Zhura} The structural transition of biphenyl (P2P) from monoclinic to triclinic has been suggested to result from the nonplanar molecules at low temperature.\cite{Charb} This phase transformation was also confirmed by Raman spectroscopy measurements.\cite{Bree} For POPs with a long C-C chain length, the ratio of the integral intensity of the 1280 cm$^{-1}$ and 1220 cm$^{-1}$ modes in the Raman spectra has often been used to distinguish the chain lengths between different compounds.\cite{Krich,Heime,Ohtsu} It thus serves as an indicator of the planarization of molecules at ambient conditions\cite{Cuff,Heime-1} or high pressures.\cite{Heime-2,Marti} Meanwhile, the features of the two peaks at around 1600 cm$^{-1}$ is also particular. Therefore, these two peaks have also been suggested to estimate the conjugation of the molecules.\cite{Ohtsu,Heime,Heime-2,Marti} The identification of the chain length effects on POPs will be more clear if the basic starting compound $-$ benzene is included. In such a way, the vibrational modes of the intra- and intermolecular terms can be easily analyzed and compared. More information can be drawn for the librational motions and C-H stretching modes at high frequencies. The Raman scattering study of POPs with a comparison with benzene is thus desirable to understand the contributions of the phenyl rings interactions.

In this work, we present Raman scattering spectra of benzene, biphenyl, $p$-terphenyl, $p$-quaterphenyl, $p$-quinquephenyl, and $p$-sexiphenyl at ambient conditions. The spectra are detected with 488 nm, 532 nm, and 660 nm excitations in a frequency range from lattice vibrations to the high frequency of the C-H vibrations. The differences among these materials are shown to clarify their different characteristics. The results are compared with previously published data.\cite{Krich,Heime,Ohtsu,Cuff,Heime-1,Heime-2,Marti} A new method is introduced to estimate the chain length of POPs and PPP materials.

\section{EXPERIMENTAL DETAILS}

In the experiments all the high-purity samples were purchased from Sinopharm Chemical Reagent and Alfa Aesarare. They were all sealed in quartz tubes with a diameter of 1-mm for Raman-scattering experiments in a glove box with moisture and oxygen levels less than 0.1 ppm. The wavelengths of the exciting laser beam were 488 nm, 532 nm, 660 nm, respectively. The power was less than 2 mW before a $\times$20 objective to avoid possible sample damage. An integration time of 20 s was used to obtain the spectra. The scattered light was focused on 1800 g/mm grating and then recorded with a 1024 pixel Princeton charge-coupled device.

\section{RESULTS AND DISCUSSION}

\subsection{Raman spectra of benzene and POPs}

\begin{figure}[htbp]
\includegraphics[width=0.48\textwidth]{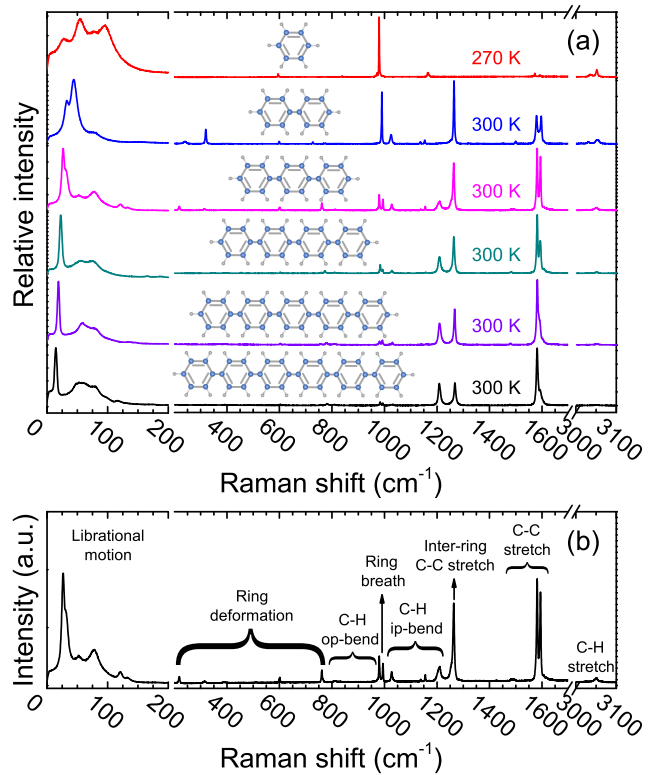}
\caption{(a) Raman spectra of benzene, biphenyl, $p$-terphenyl, $p$-quaterphenyl, $p$-quinquephenyl and $p$-sexiphenyl measured at 270 K and 300 K. The exciting wavelength is 660 nm. Curves have been broken at 200 cm$^{-1}$, 1500 cm$^{-1}$ and 3000 cm$^{-1}$ for clarity. (b) Classifications of the vibration modes of $p$-terphenyl were taken from Ref. [40].}
\end{figure}

Figure 1(a) shows the Raman spectra of benzene and the five POPs excited by a 660 nm laser, and measured under the same conditions. In order to compare these with the solid POPs, we chose the Raman spectra of crystallized benzene at 270 K. The solid-state benzene belongs to the orthorhombic system, whereas the POPs are monoclinic at ambient conditions. The spectra of compound POPs are almost identical to one another except for some differences in their relative intensities. They are also similar to benzene except in the low-frequency spectral region and the peaks corresponding to the interaction of the phenyl rings. Figure 1(b) is the classifications of the $p$-terphenyl vibration modes.\cite{Honda}

The major features in Fig. 1(a) are noted in the following points:

\begin{enumerate}
  \item The librational motion mode, the modes associated with the ring breath, the 1220 cm$^{-1}$ and 1280 cm$^{-1}$ modes, and the peaks around 1600 cm$^{-1}$ exhibit stronger intensity than other peaks.
  \item There are three peaks around 800 cm$^{-1}$. The lower frequency peak has a red-shift with increasing the chain length. The middle peak has a blue-shift, and increases the intensity as the number of the phenyl rings is increased. The higher frequency peak has a red-shift, and also increases the intensity. All these three peaks are missing in the spectra of benzene, which indicates that these behaviors result from the interaction of the phenyl rings.
  \item The intensity of the two phonon modes near 1000 cm$^{-1}$ (which are related to the ring breath) decreases from benzene to P6P. This indicates that the intramolecular interactions significantly impact these modes.
  \item The peak around 1040 cm$^{-1}$ has blue-shift, and decreases the intensity from P2P to P6P, and this peak disappears in the benzene spectra. This peak's behavior indicates that it is associated with the terminal rings.
  \item The peaks in the high-frequency region, which are associated with the C-H stretching modes, lose intensity with increasing the number of the phenyl rings. This behavior implies that these peaks are associated with the terminal rings. Otherwise, the decreased intensities of the C-H stretching modes can also result from the ratios of the H and C atoms H/C in POPs monotonously decreasing with increasing the chain length. In fact, H/C = (4n+2)/6n, where n is the number of phenyl rings. Therefore, H/C goes from 1 for n=1 (benzene) to 2/3 for n=$\infty$ (infinite chain).
\end{enumerate}

\subsection{Lattice vibrations}

\begin{figure}[tbp]
\includegraphics[width=0.48\textwidth]{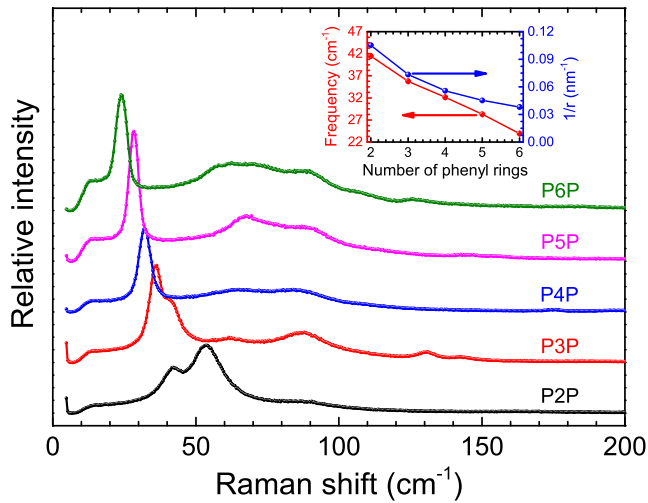}
\caption{Raman spectra of the librational motions from 0 to 200 cm$^{-1}$ of POPs, which are displaced vertically for clarity. Inset is the frequency of the first peaks and the reciprocal of $c$ of each compound.}
\end{figure}

The Raman spectra of the lattice vibrations of POPs are shown in Fig. 2. This part of the modes is associated with the structure of unit cell. The first strongest peak, which is the most important mode, is often used to detect the structural transition at extreme environments. Such as the order-disorder transition of $p$-terphenyl at low temperature\cite{Girar,Girar-1} and high pressures.\cite{Girar-2} The figure shows that the lowest frequency mode has a redshift as the number of phenyl rings is increased. The frequency fitted by a Lorentzian function is shown in the inset. However, previous XRD data showed that the length of the $c$ direction increases as the phenyl rings number increases because the long axis of the molecule is almost parallel to the $c$ direction. We compared the frequency of this mode to the reciprocal $c$ of each compound, as shown in the inset. These two curves have the same trend. Thus, we assume that this mode is associated with the vibration along the $c$ direction. The regular and obvious changes of this peak makes it a good indicator of the molecule length in PPP materials. The other peaks in the high frequencies are weak and broad, and have no obvious variation tendency, so we do not discuss them here.

\subsection{1220 cm$^{-1}$ mode and 1280 cm$^{-1}$ mode}

\begin{figure}[tbp]
\includegraphics[width=0.48\textwidth]{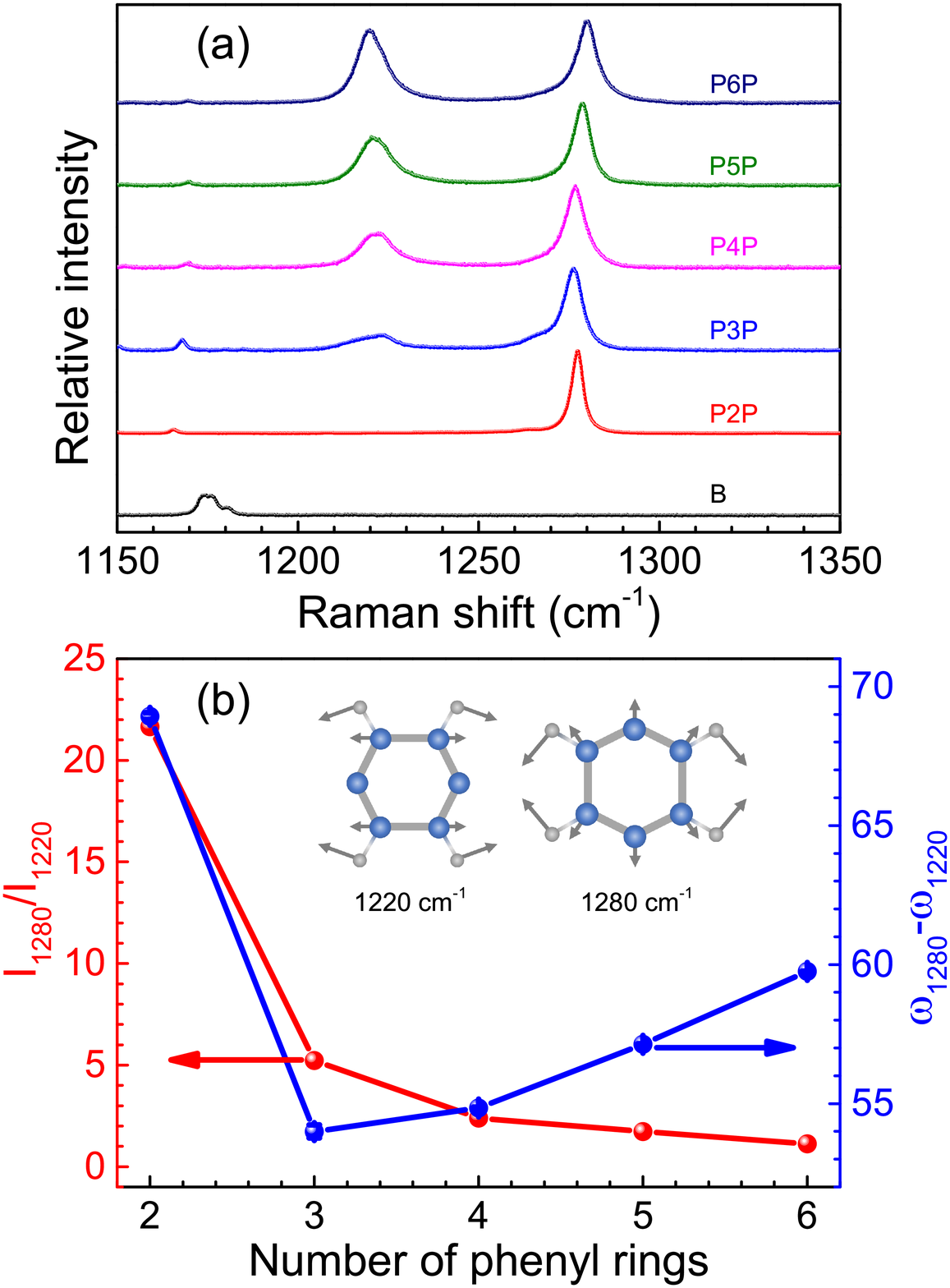}
\caption{(a) Raman spectra of the 1220 cm$^{-1}$ and 1280 cm$^{-1}$ peaks of benzene and POPs, all the spectra are normalized to the respective 1280 cm$^{-1}$ peak. (b) Ratio of the intensity of the 1280 cm$^{-1}$ mode to the 1220 cm$^{-1}$ mode and differences between them as a function of the number of phenyl rings. Inset is the intensity of each mode as a function of the number of phenyl rings.}
\end{figure}

The 1220 cm$^{-1}$ mode is associated with the C-H in-plane bend mode, and the 1280 cm$^{-1}$ mode is the C-C interring stretch mode, as shown in inset of Fig. 3(b). Figure 3(a) shows the Raman spectra of these two peaks. Both two peaks completely disappear in the Raman spectra of crystalline benzene. This reveals that these two modes result from the interaction of the phenyl rings in one molecule. Another interesting phenomenon is the intensity of these two peaks. The intensity of the 1220 cm$^{-1}$ peak increases with increasing the chain length of the POPs, whereas the 1280 cm$^{-1}$ peak has no regular changes. We fitted these two peaks with the Lorentzian function, and calculated the ratio of the intensity of these two peaks. The results are summarized in Fig. 3(b). This relative intensity ratio decreases monotonously with the increase in the chain length. Consequently, it is often used to estimate the length of PPP materials.\cite{Krich,Heime,Ohtsu}

\begin{figure}[tbp]
\includegraphics[width=0.48\textwidth]{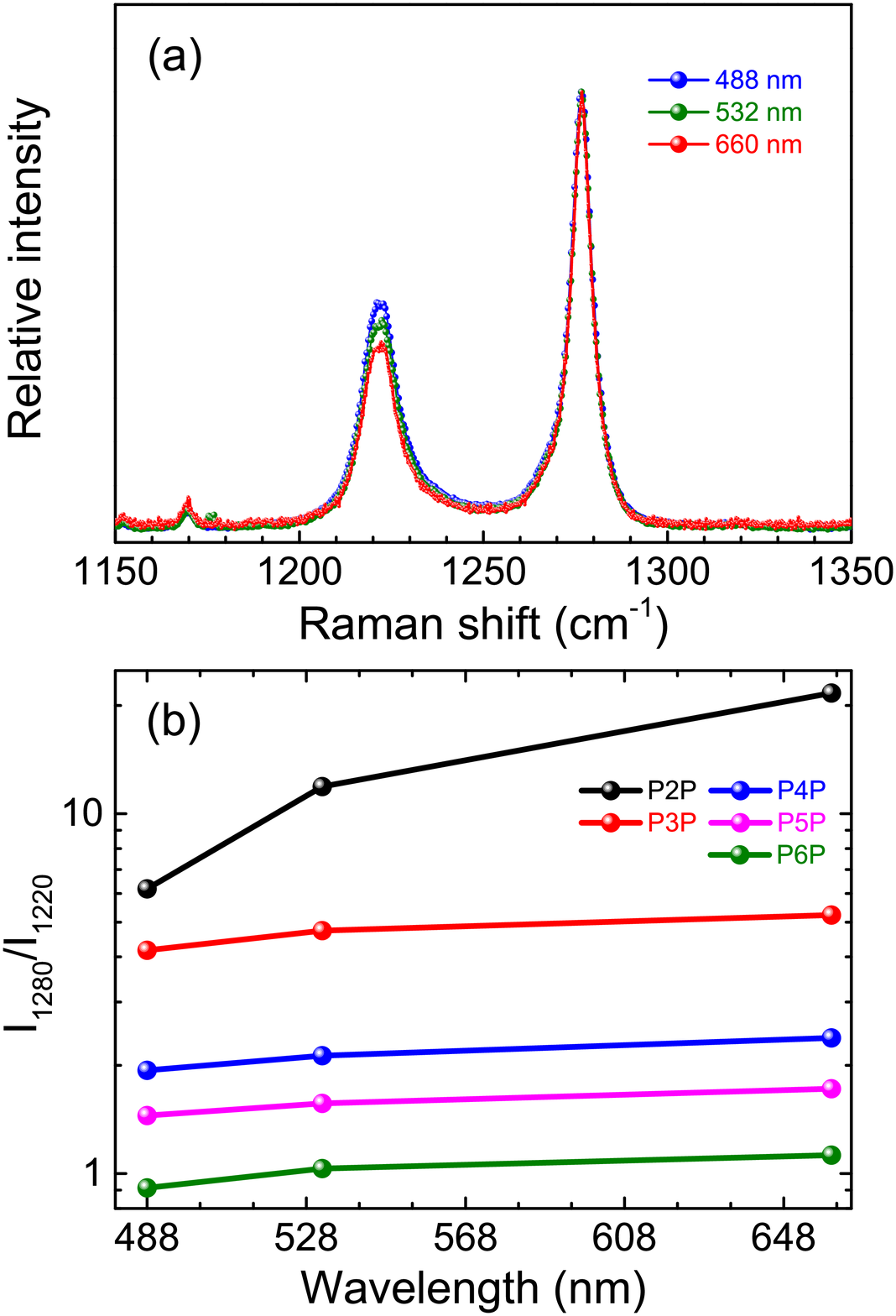}
\caption{(a) The 488 nm, 532 nm and 660 nm excited Raman spectra of the spectral region of 1220 cm$^{-1}$ and 1280 cm$^{-1}$ modes of $q$-quaterphenyl, all the spectra are normalized to the respective 1280 cm$^{-1}$ peak. (b) I$_{1280}$/I$_{1220}$ of POPs as a function of wavelength.}
\end{figure}

The ratio of the intensities of these two peaks is also an indicator of the inter-ring torsion angle.\cite{Cuff,Heime-1,Heime-2,Marti} In PPP materials, all the molecular orbitals can be divided into delocalized orbitals and localized orbitals. Theoretical calculations\cite{Heime-1,Pusch} further showed that the delocalized orbitals are mainly associated with the C atoms on the long molecular axis and strongly sensitive to the conjugation (chain length and planarity), whereas the localized orbitals are mainly contributed by the off-axis C atoms and are far less susceptible to conjugation effects. In addition, the 1220 cm$^{-1}$ mode is interpreted to be more closely related to the delocalized state than the localized state, and the 1280 cm$^{-1}$ mode is the opposite. Thus, the former mode is sensitive to conjugation, whereas the latter is rather insensitive to the chain length. As a result, I$_{1280}$/I$_{1220}$ decreases upon increasing the length of the molecules. The result exhibited in the Fig. 3(b) is consistent with the calculations.

Figure 3(b) also presents the difference of the frequency between the 1280 cm$^{-1}$ mode and 1220 cm$^{-1}$ mode as a function of the number of phenyl rings. This difference decreases from P2P to P3P, then increases from P3P to P6P. Previous literature\cite{Marti} shows that the difference in frequency between these two modes also indicates planarization. Hence, we assume that the torsion angle of P2P is abnormal compared with the other POPs, $i.e.$, the molecular interactions between the intra- and intermolecular in POPs where the molecules contain more than two phenyl rings are different from the P2P.

\begin{figure}[tbp]
\includegraphics[width=0.48\textwidth]{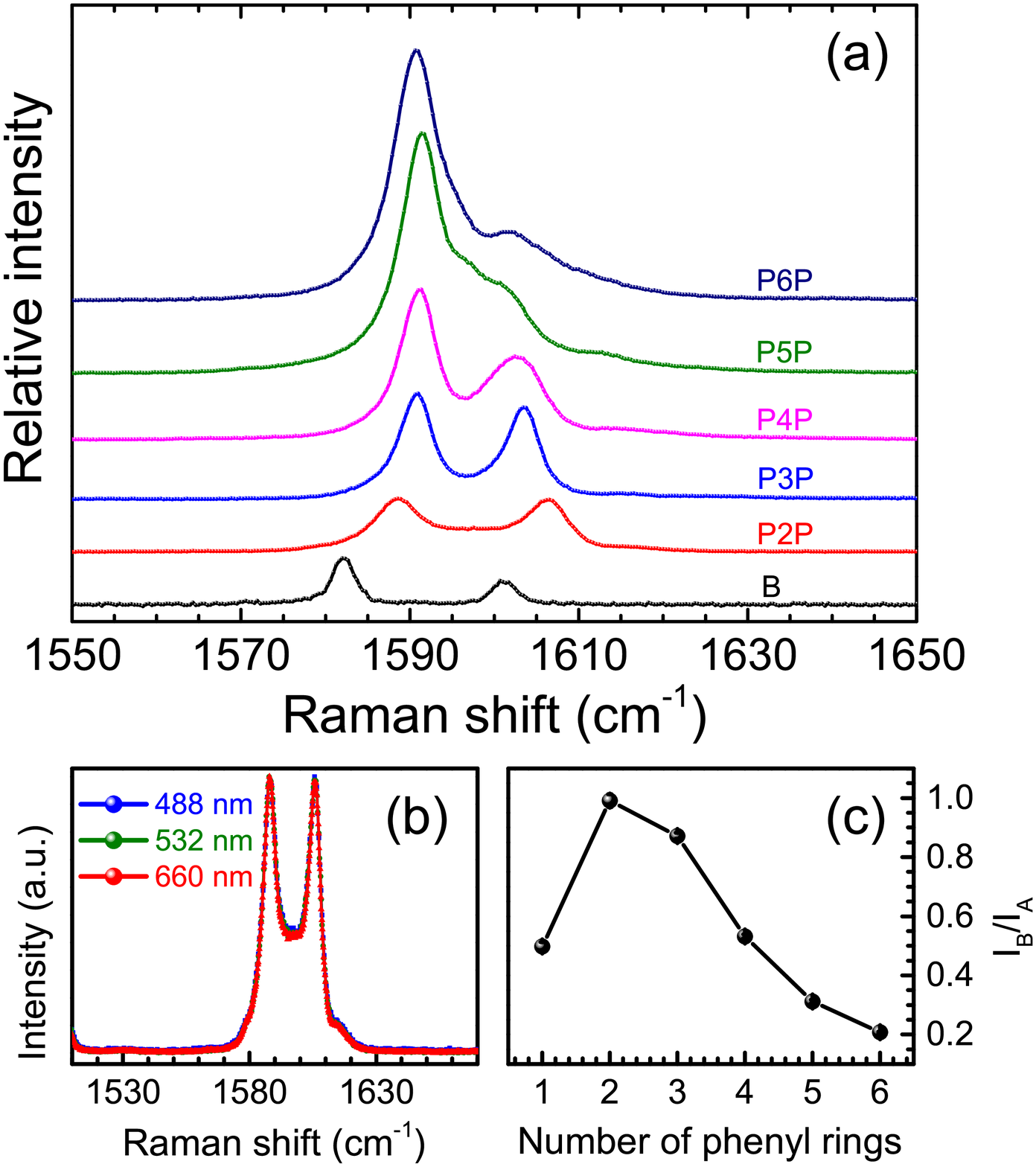}
\caption{(a) Raman spectra of the spectral region of 1600 cm$^{-1}$ modes of benzene and POPs. (b) The 488 nm, 532 nm and 660 nm excited Raman spectra of the spectral region of 1600 cm$^{-1}$ modes of biphenyl, each spectrum has normalized to the low energy peak. (c) The ratio of the intensity of high energy peak to low energy peak as a function of number of phenyl rings.}
\end{figure}

Figure 4(a) shows the Raman spectra of P4P excited by 488 nm, 532 nm, and 660 nm lasers. It is obvious that the intensity of the 1220 cm$^{-1}$ mode is closely related to the energy of the excited lasers. The 1220 cm$^{-1}$ peak increases its intensity with increasing the excited energy. However, the 1280 cm$^{-1}$ mode is insensitive to the excited energy. Thus, the I$_{1280}$/I$_{1220}$ ratio decreases as the excitation wavelength decreases. This also shows that the 1220 cm$^{-1}$ mode, which is associated with the delocalized orbitals, is more strongly pre-resonance enhanced than the 1280 cm$^{-1}$ mode as the exciting laser approaches resonance with the lower energy delocalized band.\cite{Heime-1} In Fig. 4(b), we show the ratio of the intensities of the 1220 cm$^{-1}$ mode and 1280 cm$^{-1}$ mode for POPs containing between three and six repeat units as a function of the laser wavelength. All the I$_{1280}$/I$_{1220}$ ratios of each of the POPs increase with increasing the laser wavelength. In addition, the ratio at a fixed wavelength decreases upon increasing the oligomer length.

\begin{table}[tbp]
\centering
\caption{Frequencies (in cm$^{-1}$) of $\upsilon_6$, $\upsilon_1$, and $\upsilon_8$ of benzene, biphenyl, $p$-terphenyl, $p$-quaterphenyl, $p$-quinquephenyl, and $p$-sexiphenyl.}
\begin{ruledtabular}
\renewcommand\arraystretch{1.5}
\begin{tabular}{ccccccc}
 \textbf{Compound} & \textbf{B} &  \textbf{P2P} &  \textbf{P3P} &  \textbf{P4P} &  \textbf{P5P} &  \textbf{P6P} \\\hline
  $\upsilon_6$ & 604 & 608 & 610 & 610 & 613 & 612 \\
  $\upsilon_1$ & 989 & 999 & 989 & 989 & 990 & 993 \\
  $\upsilon_8$ & 1600 & 1606 & 1603 & 1603 & 1600 & 1601 \\
\end{tabular}
\end{ruledtabular}
\end{table}

\subsection{Modes around 1600 cm$^{-1}$ }

The $E_{2g}$ modes $\nu_8$ at around 1600 cm$^{-1}$ are associated with the C-C stretch. The Raman spectra around 1600 cm$^{-1}$ from benzene to P6P are shown in Fig. 5(a). Two peaks appear in this region, except the P5P and P6P, which have an obscure peak in the middle of these two peaks. A great deal of work has focused on these two peaks as their prominent features are related to the oligomer length and planarity of the molecules.\cite{Ohtsu,Heime-3,Marti} Previous works on benzene\cite{Angus-1,Wilso,Tto} observed the 1600 cm$^{-1}$ doublet. The benzene line doublet is suggested to result from the "resonance splitting" to a fundamental band (low energy peak) with a frequency close to 1596 cm$^{-1}$ and a combination tone (high-energy peak) from the $E_{2g}$ fundamental $\nu_1$ at 992 cm$^{-1}$ and the $E_{2g}$ fundamental $\nu_6$ at 606 cm$^{-1}$. This idea is confirmed by the experimental results. Later, this idea was applied to the oligomer materials, which agreed with the experimental results.\cite{Heime-3} The frequencies of the $\nu6$, $\nu_1$, and $\nu_8$ are summarized in Table \uppercase\expandafter{\romannumeral1}. Our data is totally consistent with this theory. Figure 5(b) is the Raman spectra of biphenyl excited by 488 nm, 532 nm, and 660 nm lasers. Unlike the 1220 cm$^{-1}$ and 1280cm$^{-1}$ peaks, excitations of different energy have nearly no effect on this 1600 cm$^{-1}$ doublet.

Then again, the intensities of these two peaks also exhibit a regular change like the 1220 cm$^{-1}$ and 1280cm$^{-1}$ peaks. We calculated the intensity ratio I$_B$/I$_A$ of the high-energy peak (B) and the low-energy peak (A). The results are shown in Fig. 5(c). The ratio of I$_B$/I$_A$ rockets from benzene to P2P, then monotonous decreases with increasing the oligomer length. The trend of the ratio to the long-chain polymer is almost zero, this is consistent with the Raman spectra of PPP,\cite{Krich-1,Cuff} in which no double peaks were observed. Thus, I$_B$/I$_A$ is also a pretty good indicator for estimating the length of the POP molecules. The abnormal tendency of benzene is assumed to result from the interaction among the phenyl rings in POPs. The same phenomenon can also be observed from the Raman spectrum of benzene, which has the highest intensity of $\nu_1$ mode in these six materials but the combination band is the weakest peak.

\subsection{Discussion}

The estimation of the chain length in PPP materials is fundamental but crucial for exploring novel properties, such as higher conductivity, superconductivity, thermoelectricity, etc.
To data, the indicators of molecular length are summarized as follows: a) the ratio of the intensity of the 1220 cm$^{-1}$ and 1280 cm$^{-1}$ phonon modes, and b) the ratio of the intensity of the two phonon modes around 1600 cm$^{-1}$. These two indicators all have regular changes depending on the chain length. Thus, they can be used to estimate the chain length. However, these two indicators are not always accurate. The former is almost a constant for the PPP materials with a long chain length [see Fig. 3(b)], and all these peaks exhibit splits in low temperature.\cite{Heime-2,Barba} Thus, their ratios will be seriously influenced, $i.e.$ the conjugation cannot be exactly determined. On the other hand, the Fermi resonance to the excited laser, the inter-ring torsion angles, high-pressure conditions, and so on, also impact the ratios.

Another indicator is the lowest frequency peak, which belongs to the lattice mode. The frequency of this peak has a monotonous change. It is insensitive to the temperature\cite{Girar,Girar-1} and pressure,\cite{Girar-2} and cannot resonate with the excited laser due to the low energy. For these reasons, the frequency of this mode is a better indicator to estimate the chain length of molecules.

\section{Conclusions}

We have reported Raman scattering spectra of benzene, biphenyl, $p$-terphenyl, $p$-quaterphenyl, $p$-quinquephenyl, and $p$-sexiphenyl. A detailed comparison between $p$-oligophenyls and benzene has been done to analyze the vibrational modes. Indeed, the phonon mode at around 1200 cm$^{-1}$ is strongly influenced by the chain length of the molecules, whereas the 1280 cm$^{-1}$ phonon band is far less susceptible to the length. Thus, the ratio of their intensities is observed to have a systematic variation with the chain length of PPP materials at ambient conditions. However, the difference of the frequencies between these two modes has an anomaly at P2P. This is because the interaction of the phenyl rings is different between P2P and POPs containing more than two phenyl rings. Meanwhile, their intensities are also sensitive to the excited lasers, decreasing with increasing the energy. These features indicate that the intensity ratio of the 1220 cm$^{-1}$ and 1280 cm$^{-1}$ modes is not a good indicator of the chain-length dependence of PPP materials. The phonon modes with lower energy and higher energy around 1600 cm$^{-1}$, which  respectively correspond to a fundamental band and a combination band  from the $\nu_1$ mode at 992 cm$^{-1}$ and the $\nu_6$ mode at 606 cm$^{-1}$, have been examined in benzene and $p$-oligophenyls. Although the ratio of their intensities exhibits a monotonous change with an increase in the chain length, this tendency is not obeyed by adding benzene. This anomaly is attributed to the interaction of the phenyls rings. Compared to the above well-established behaviours, we have found that the frequency of the first librational motion mode exhibits a monotonous change with increasing the chain length. This behavior is assumed to associated with the vibration along the $c$ direction of the unit cell because of the same trend between the frequency of this phonon mode and the reciprocal $c$ axis. Such an obvious and ordered behaviour of this phonon mode makes it a better indicator of the molecule chain length than others.

We thank Freyja O'Toole for the language editing.

\end{document}